\documentstyle[aps,prd]{revtex}
\begin{document}
\draft
\preprint{IUCAA xxx/1995}
\title {Estimation of parameters of gravitational waves from coalescing
binaries}
\author{R. Balasubramanian \and B.S. Sathyaprakash \and  S. V. Dhurandhar}
\address{Inter-University Centre for Astronomy and Astrophysics, \\
Post Bag 4, Ganeshkhind, Pune 411 007, India}
\date{\today}
\maketitle
\begin{abstract}
In this paper we deal with the measurement of the parameters of the
gravitational
wave signal emitted by a coalescing binary signal. We present the results of
Monte
Carlo simulations carried out for the case of the initial LIGO, incorporating
the first
post-Newtonian corrections into the waveform. Using the parameters so
determined,
 we estimate the direction  to the source. We stress the use of the
time-of-coalescence rather than the time-of-arrival of the signal to determine
the direction of the source. We show that this can considerably reduce the
errors in the
determination of the direction of the source.
\end{abstract}

\vspace{.5in}

A major source for the planned interferometric gravitational wave detectors
like LIGO~\cite{LIGO} and VIRGO~\cite{VIRGO} is the radiation emitted by a
binary system of
stars during the last few minutes of its evolution just before the two stars
coalesce.
The interferometers are expected to have high enough sensitivity to detect
sources at cosmological
distances. Since the waveform emitted during a binary coalescence can be
modeled fairly
accurately it is possible to enhance the visibility or
the `signal-to-noise' ratio\footnote{henceforth denoted as SNR},
(defined later in the text in eq. \ref{rhodef})
of the signal in noisy data by employing
such powerful techniques such as Weiner filtering \cite{Th300,SCH89}. With the
aid of such methods one
essentially compares signal energy to that of the noise as opposed to the case
of a burst signal where one has to contend with the signal power vis-a-vis
noise power
\cite{SCH89}.
 The fact that the signal waveform can be predicted very well also implies that
the parameters
of the waveform can be estimated to a high accuracy. For instance, by tracking
the radiation
emitted by a neutron star-neutron star\footnote{In this paper a neutron star is
considered
to have a mass of $1.4~M_\odot$,and a black hole
will be considered to have a mass of $10.~M_\odot$} (NS-NS)  binary starting
from
$10$~Hz. to say $1$kHz., the masses can be determined to better than a few
percent
\cite{FC93,CF94,KKS95,JKKT,BSD95,Mark}.
Observation of several coalescence events could be potentially used to
determine Hubble parameter to
an accuracy of $\sim 10$\% \cite{PW95}.
Once in a while, when an event produces a very high SNR
$(\ge 40)$
it would be possible to infer the presence of gravitational wave tails in the
signal waveform and test general relativity in the strongly non-linear regime
\cite{BS95,BS94}. These are, but a modest list of interesting physical and
astrophysical
information that the observation of gravitational waves is expected to bring
forth.

In the recent past the problem of optimal detection and estimation of
parameters of the binary
has gained considerable attention, thanks to the funding of the American LIGO
and
Franco-Italian VIRGO projects.
The chirp waveform is expected
to be detected with the aid of Weiner filtering, wherein the detector output is
correlated with a set
of templates which together span the relevant range of the parameters of the
binary waveform.
When a signal
is present in the detector output, the filter matched to that particular signal
will on the
average obtain the largest possible correlation with the detector output
amongst all other filters:
thus enabling detection as well as estimation of parameters. The parameters of
the template that obtains
the maximum correlation is an un biased estimate of the actual signal
parameters.
Of course, the detector output also contains noise which can effect the
correlation thereby giving rise to spurious maxima even when the parameters of
the template and those of the signal are mismatched. Thus,
the measured parameters are in error. These errors tend to reduce with
increasing SNR.

Based on analytical computation of the covariance matrix of the errors in the
estimation of parameters and the covariances among them (for the advanced
LIGO),
it is now generally agreed that the chirp mass  (which is a combination of the
masses of the binary stars)
can be determined at a relative accuracy $\sim$ .1\%-1\% at a SNR of 10.
The reduced mass, as it turns out, will have a much larger error especially if
the spins are large.
In the computation of the covariance matrix one makes the crucial assumption of
high SNR. As a result the covariance matrix only gives a bound on the errors in
the estimation of parameters and a more rigorous or a different method of
computing the variances and covariances is in order.
Alternatively one can determine the errors through the Monte Carlo simulation
method.
The basic idea here is to mimic the actual detection problem on a
computer by adding numerous realizations of the noise to the signal and then
passing the resultant time
series through a bank of Weiner filters each matched to a gravitational wave
signal from a
coalescing binary system with a particular set of parameters. The covariance
matrix can be computed by using the ensemble of measured values of the signal
parameters. Such a method is robust in the sense that
it does not make any assumptions about the SNR and therefore is much closer to
the actual detection problem. It also squarely addresses the difficulties faced
in a realistic data analysis exercise
such as the finite sampling rate, the finiteness tof spacing in between the
filters, {\em etc}.
For details on the Monte Carlo method see Balasubramanian, Sathyaprakash and
Dhurandhar \cite{BSD95}.

	In this paper we summarize the outcome, of the first in a series, of Monte
Carlo simulations which suggests that the covariance matrix grossly
underestimates the errors in the estimation of the parmeters
by over a factor of 3 at an SNR of 10.

	The standard method to deduce the direction to the direction to the source of
the
gravitational radiation is to use the differences in the times-of-arrival in a
network of
three or more detectors. We point out that since the instant of coalescence of
a binary system can
be measured to a much greater accuracy than that of the time-of-arrival,
 the former can be employed to infer the direction of the source at a much
lower uncertainty
than has been so far thought.

	In the point mass approximation the restricted first-order post-Newtonian
gravitational radiation
emitted by a binary system of stars induces a strain in the detector which can
be written as
\begin{equation}
\label{hn}
h(t) = A (\pi f(t) )^{2/3} \cos \left [\varphi (t) \right ],
\end{equation}
where $f(t)$ is the instantaneous gravitational wave frequency, which at this
level of approximation
is equal to twice the orbital frequency and is implicitly given by:
\begin{equation}
\label{taus}
t - t_a = \tau_0
\left [ 1 - \left ( {f \over f_a} \right )^{-8/3} \right ] +
\tau_1\left[1 - \left(\frac{f}{f_a}\right)^{-2}\right]
\end{equation}
and $\varphi(t)$ is the phase of the signal given by
\begin{equation}
\label{pha}
\varphi (t) = {16 \pi f_a \tau_0 \over 5} \left [ 1 - \left ({f\over f_a}\right
)^{-5/3} \right]
+ 4 \pi f_a \tau_1 \left [ 1 - \left ( {f\over f_a} \right )^{-1} \right ] +
\Phi
\end{equation}
where $t_a$ and $\Phi$ are, respectively,
 the so called time-of-arrival and phase  at which the signal reaches a
frequency $f_a$;
$\Phi$ is a constant depending on the relative orientations of the binary orbit
and the arms of the interferometer; $\tau_0$ and $\tau_1$ are constants, having
dimensions of time, depending on the masses
$m_1$ and $m_2$ of the binary system. They are referred to as chirp times and
are given by,
\begin{eqnarray}
\label{tausd}
\tau_0 &=& {5 \over 256} {\cal M}^{-5/3} (\pi f_a)^{-8/3},\\
\tau_1 &=& {5 \over 192\mu (\pi f_a)^2} \left ({743\over 336} + {11\over 4}
\eta\right),
\end{eqnarray}
where $\mu$ is the reduced mass, $\eta$ is  the reduced mass divided by the
total mass $M$,
and ${\cal M} = \mu^{3/5}M^{2/5}$ is the chirp mass.
In this level of approximation the post-Newtonian waveform is thus
characterised by a set of five parameters: The amplitude parameter $A$, the
time-of-arrival $t_a$, the phase at the time of arrival
$\Phi$, and the two masses $m_1$ and $m_2$. We shall find below a more
convenient parameterization of the
waveform.

The Fourier transform of the waveform in the stationary phase approximation is
given by
\begin{equation}
\label{sta}
\tilde h (f) = {\cal A} f^{-7/6} \exp \left
[i\sum_{\mu=1}^4\psi_\mu(f)\lambda^\mu -i{\pi\over 4}\right]
\end{equation}
where $\cal A$ is a normalization constant and is fixed by specifying the SNR,
 and $\psi_\mu(f)$ are functions only of frequency given by,
\begin{eqnarray}
\label{psis}
\psi_1 & = & 2\pi f, \\
\psi_2 & = & -1, \\
\psi_3 & = & 2 \pi f  -{ 16 \pi f_a \over 5}+ {6\pi f_a \over 5}
\left ( {f\over f_a} \right )^{-5/3},\\
\psi_4 & = & 2\pi f - 4\pi f_a + 2\pi f_a \left ({f\over f_a}\right)^{-1}.
\end{eqnarray}
Here $\lambda^\mu$ are related to the set of parameters introduced earlier and
are given by,
\begin{equation}
\lambda^0 = {\cal A} \ \mbox{ and }\  \lambda^\mu =\{t_C,~\Phi_C,~{\cal
M},~\eta\},\ \mu = 1,\ldots,4 \ ,
\label{lambdas}
\end{equation}
where,
\begin{eqnarray}
\label{phc}
 t_C &=& t_a + \tau_0 + \tau_1  \\
\Phi_C &=& \Phi - 2\pi f_a t_a + \frac{16\pi f_a\tau_0}{5} + 4\pi f_a\tau_1
\end{eqnarray}
For $f < 0$, the Fourier transform can be obtained by the relation $\tilde
h(-f) = \tilde h^*(f)$,
obeyed by all real functions $h(t)$.

Central to the computation of the covariance matrix is the scalar product
defined over the space of the signal waveforms. Given waveforms
$h(t;\bbox{\lambda})$ and $g(t;\bbox{\lambda})$ their scalar product is defined
by,
\begin{equation}
\label{scal}
\left\langle h,g\right\rangle = \int\limits_{f_a}^\infty\frac{\tilde
h(f;\lambda^\mu)\tilde g^*(f;\lambda^\mu)}{S_n(f)}df + \mbox{c.c.}
\end{equation}
where $S_n(f)$ is the two sided detector noise power spectral density.
The scalar product defines a norm on the vector space.
The SNR can now be defined for a signal $h(t;\bbox{\lambda})$
when it is passed through the matched filter as the norm of the signal {\em
i.e.},
\begin{equation}
\label{rhodef}
\rho = \left\langle h,h\right\rangle^{1/2}.
\end{equation}
 The covariance matrix $C_{\mu\nu}$
is the inverse of the Fisher information matrix $\Gamma_{\mu\nu}$ given by,
\begin{equation}
\label{Gamma}
\Gamma_{\mu\nu} = \left < {\partial {\bf s} \over \partial \lambda^\mu},
{\partial {\bf s} \over \partial \lambda^\nu} \right >;
\ \ \ C_{\mu\nu} = \Gamma^{-1}{\mu\nu}.
\end{equation}
The diagonal elements of the covariance matrix provide us an estimate of the
variances to be expected in the measured values of the parameters in a given
experiment. In Fig. \ref{fig_1} we have shown the behaviour of the $1\sigma$
uncertainties (square root of the variances)
in the estimation of parameters $t_a$,$\tau_0$,$\tau_1$ and $\tau_C$, as a
function of the SNR.
As is well known the covariance matrix predicts that the errors fall off in
inverse propotion
to the SNR. It is to be noted thet this is valid only in the `high' SNR limit.

In the actual detection problem the detector output $x(t)$ is filtered through
a host of search templates
corresponding to {\em test} parameters,$_t\bbox{\lambda}$ and the template that
obtains the maximum correlation with the output
gives us the measured values $_m\bbox{\lambda}$ of the signal. These measured
values will in general
differ from the actual signal parameters. With the aid of a large number of
detectors
one obtains an ensemble of measured values $_m\bbox{\lambda}$.  The average of
such an ensemble provides
provides us with an estimate $_e\bbox{\lambda}$ and the variances and
covariances $\sigma_\lambda$
computed using  $_m\bbox{\lambda}$  helps us in deducing the errors in the
estimation
and how the different parameters are correlated:
\begin{equation}
\label{sigmas}
{{\rm _e}{\bbox {\lambda}}} = \overline{{\rm _m}{\bbox {\lambda}}},\ \ \ \ \
\sigma_{\bbox {\lambda}}^2 = \overline{ \left ( { {{\rm _m}{\bbox {\lambda}}} }
-
\overline{ {{\rm _m}{\bbox {\lambda}}}} \right )^2},  \ \ \ \
 \Sigma^{\mu\nu} = \overline{ {{\rm _m}\lambda}^\mu \ {{\rm _m}\lambda}^\nu
\over \sigma_\mu \sigma_\nu},
\end{equation}
where an overbar denotes an average over an ensemble and $\Sigma_{\mu\nu}$ the
correlation coefficients.
In reality we will have only a few detectors and hence a numerical simulation
needs to be carried out to
deduce the errors in the estimation of parameters.
 We have carried out such a numerical simulation by using in excess of 5000
realizations of detector noise. Thus, the results of our simulations are
statistically significant, the errors in the estimation of various statistical
quantities such as the mean and the variance are negligible
and hence we do not plot the error bars in our curves.
In Figure \ref{fig_1} the dotted line corresponds to the behaviour of the
errors in the estimation of various parameters computed at several values of
the SNR. At low values of the SNR $(\rho \sim 10)$ there
is a significant departure of the observed errors from those predicted by the
covariance matrix.
However at an SNR of $\ge 30$ the two curves merge together indicating the
validity of the covariance
matrix at high enough SNRs. Since most of the events which the interferometric
detectors will observe are
expected to have an SNR less than $10$, we conclude that the accuracy in the
determination of the parameters
is not as high as was thought to be, based on the values of the covariance
matrix. Detailed analysis suggests that this discrepancy is larger when higher
post-Newtonian corrections are taken into account. Consequently
a more exhaustive analysis than has been reported here or elsewhere
\cite{BSD95} is in order.
We are in the process of carrying out simulations by taking a second order
post-Newtonian waveform and including other physical effects such as the
eccentricity of the binary \cite{Mark}.

With reference to Figure \ref{fig_1} we point out that the instant of
coalescence $t_C$ can be determined
to an accuracy much better than the time-of-arrival $t_a$. Typically
$\sigma_{t_C}$ is a factor
of 50 less than  $\sigma_{t_a}$. Consequently with the aid of $t_C$ we can fix
the direction to
 the source at a much higher accuracy than with $t_a$ as can be seen from
equation (\ref{phc}),
$t_C$ is the sum of $\tau_0$, $\tau_1$ and $t_a$, and the errors in these
parameters tend to cancel
because of the presence of negative covariances.
 It is to be noted, however, that as of now we do not
know the orbit of the binary accurately enough, to predict the exact instant of
the coalescence.
In fact the frequency cutoff imposed by the onset of the plunge orbit will make
it
difficult to calculate $t_C$. Moreover, for a realistic detector the noise
increases with the
frequency beyond about $150$~Hz. and the frequency which will be of more
interest to us is the one where
the power spectrum of the signal divided by the power spectrum of the noise in
the
detector reaches a maximum. This of course assumes that all the detectors to
used for direction
measurement are identical.

Let $t$ denote a convenient time parameter which if measured at three detectors
gives the
direction to the source. For our purpose we take this parameter to be either
$t_C$ or $t_a$.
We will assume a very simple configuration  of three identical detectors placed
at the
vertices of an equilateral triangle on the equitorial plane of the earth. We
will take the separation
between any two detectors as $L$ and consequently the maximum time delay
induced by $L$ as
$\Delta = L/c$ where $c$ is the velocity of light.
Figure \ref{fig_2}
illustrates such a network and the coordinate system used. The $X$ axis passes
through the
detectors labelled 1 and 2 and has its origin at the former. The positive $Y$
axis is chosen as
shown and the $Z$ axis is perpendicular to this plane and forms a right handed
coordinate system
with the other axes.
Given the  values $t_1$, $t_2$ and $t_3$ as the measured values of $t$ in the
three detectors
respectively we can deduce the direction to the source via the time delays
$\gamma =
t_2 - t_1$ and $\delta = t_3 - t_1$. As the noise in each detector is
uncorrelated with the noise
in the others $t_1, t_2$ and $t_3$ are uncorrelated and is we assume that all
the detectors are
identical then the errors in $t$ in all the detectors will be the same.
However, $\gamma$ and $\delta$
will have non-zero covariances. Thus,
\begin{equation}
\label{sigmats}
\sigma_{t_1} = \sigma_{t_2} = \sigma_{t_3} = \sigma;\ \ \ \
\sigma_{\gamma} = \sigma_{\delta} = \sqrt{2}\sigma.
\end{equation}
The angles $\phi$ and $\theta$ are related to $\gamma$ and $\delta$ by,
\begin{equation}
\label{angsp}
\phi = \tan^{-1}\left[\frac{2\delta - \gamma}{\sqrt{3}\gamma}\right],
\end{equation}
and,
\begin{equation}
\label{angst}
\theta = \sin^{-1}\left[\frac{2 \gamma \sqrt{\gamma^2 + \delta^2 -
\gamma\delta}}
{\Delta\left(2\delta - \gamma\right)}\right].
\end{equation}
The errors in the measurement of the time parameter will induce errors in the
measurement of the
angles. Assuming the errors in the time parameter to be small we can write the
expressions for
$\sigma_\phi$ and $\sigma_\theta$ as,
\begin{equation}
\label{angspe}
\sigma_\phi = \frac{\sigma}{\Delta} g_\phi(\theta,\phi) \ \ \  \mbox{ and } \ \
\
\sigma_\theta = \frac{\sigma}{\Delta} g_\theta(\theta,\phi)\ ,
\end{equation}
where,
\begin{equation}
\label{angste}
g_\phi(\theta,\phi) = \frac{1}{\sin\theta} \ \ \ \mbox{ and }\ \ \
g_\theta(\theta,\phi) = \sqrt{\frac{\cos^4\phi - \cos^2\phi - 1}
{-1 + 3\cos^2\phi - 2\cos^4\phi - \cos^2\phi\cos^2\theta\sin^2\theta} } .
\end{equation}
The factors $g_\phi(\theta,\phi)$ and $g_\theta(\theta,\phi)$ are typically of
order unity
for most directions.
For earth based detectors the value of $\Delta$ is $\approx 15$~ms. This
certainly rules
out the use of $t_a$ to determine the direction as the error in this parameter
even
for high SNRs is much more than $\Delta$. Though for the initial LIGO we can
use $t_C$,
the errors at an SNR of 10 will be around two degrees which is too large to
make an optical
identification of the source. In order to determine the direction to arcsecond
resolution
one needs the sensitivity of the advanced LIGO.

We would like to end this paper with the remark that though  the inclusion of
higher
order post-Newtonian corrections is expected to bring down the precision with
which the
astrophysical parameters can be determined, they are not a cause for worry for
the
detection problem. Even at the second post-Newtonian correction the signal can
essentially be detected
with the aid of a one-dimensional lattice of templates \cite{BSD95}. However,
more extensive numerical
Monte Carlo simulations need to be performed to gain further insight into the
estimation problem.

\acknowledgments
The authors would like to thank Biplab Bhawal
for useful discussions. R.B. would like to acknowledge  CSIR, INDIA for
financial support through the Senior Research Fellowship.

\begin{figure}
\caption{Dependence of the errors in the estimation of parameters of the
post-Newtonian waveform
{\em i.e.} \{$\sigma_{\tau_0},\sigma_{\tau_1},\sigma_{t_a},\sigma_{t_C}$\}
as a function of SNR. The solid line
represents the theoretically computed errors whereas the dotted line represents
the errors obtained
through Monte Carlo simulations. The simulations have been carried out for case
of a
$10M_\odot-1.4M_\odot$ binary system. The errors in the parameters are
expressed in ms.}
\label{fig_1}
\end{figure}

\begin{figure}
\caption{This figure  illustrates the network of three detectors and the choice
of the
coordinate system employed. The $Z$ axis is perpendicular to the plane of the
paper and points
upward. The angle $\theta$ is defined as the angle which the direction vector
makes with the
positive $Z$ axis and $\phi$ is defined on the $X-Y$ plane as shown in the
figure.
}
\label{fig_2}
\end{figure}

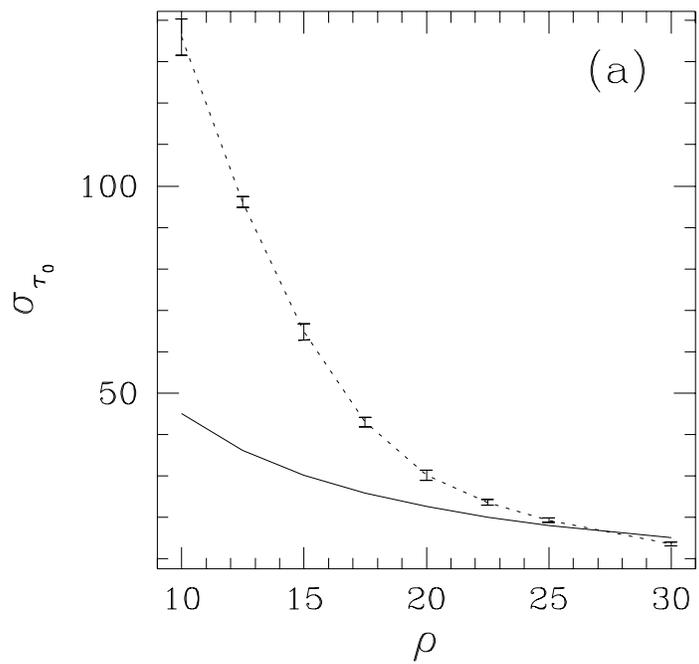
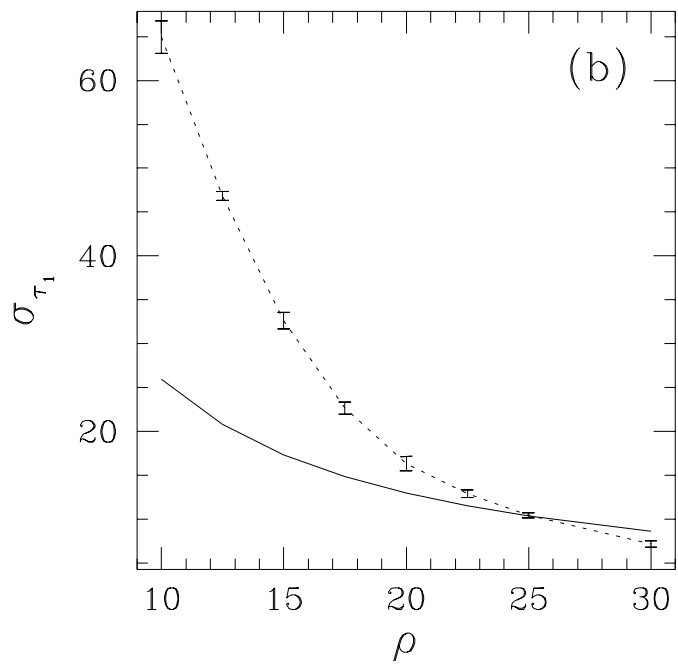
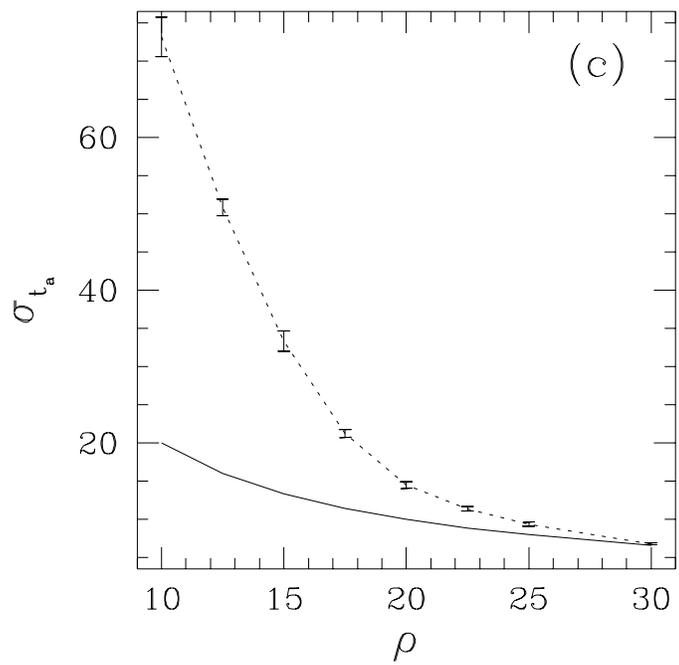
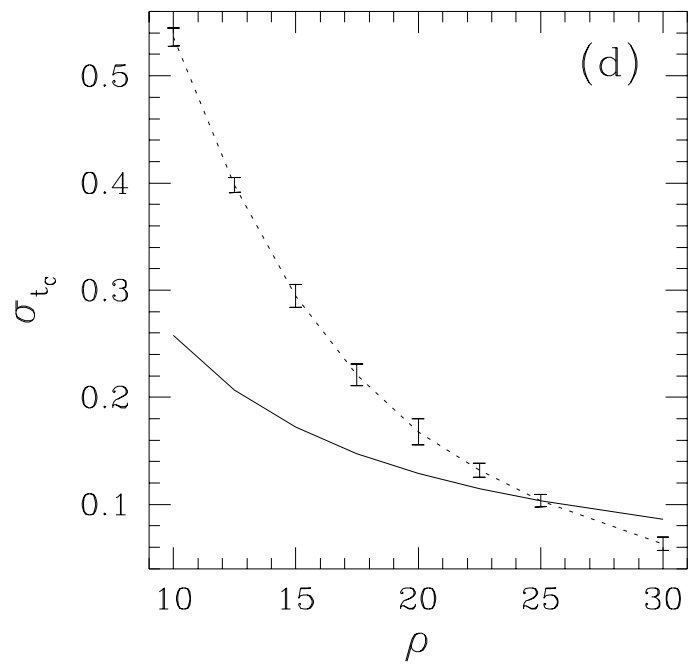
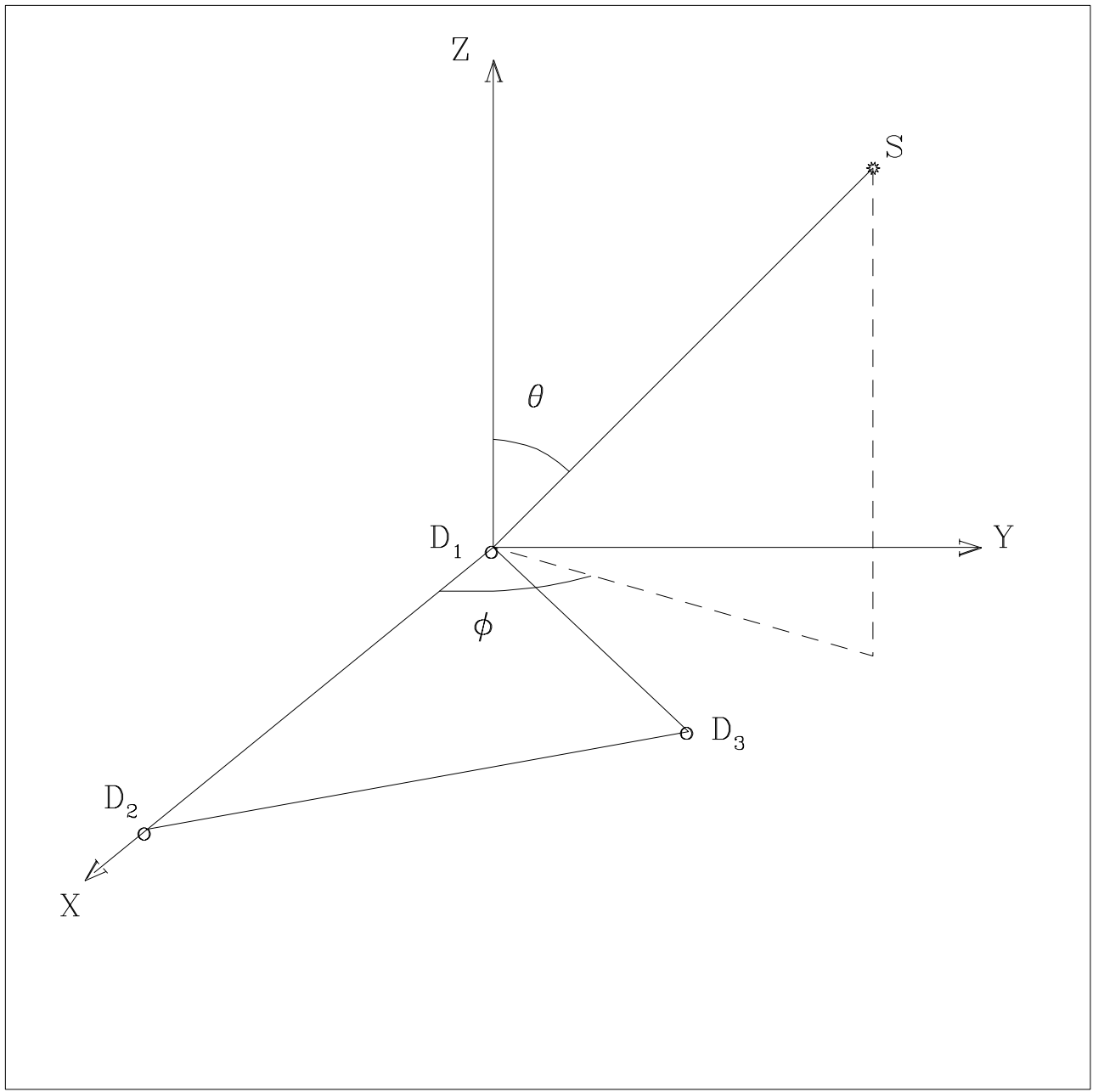
\end{document}